\documentclass[prl,twocolumn,nopacs,nofootinbib]{revtex4-2}

\usepackage{amsmath}
\usepackage[dvipsnames]{xcolor}
\usepackage{nicefrac}
\usepackage{graphicx}
\usepackage{microtype}
\usepackage{physics}
\usepackage{mathrsfs}
\usepackage{bbold}

\usepackage[english]{babel}
\usepackage[applemac]{inputenc}
\usepackage[T1]{fontenc}

\newcommand{\beq}{\begin{equation}}
\newcommand{\eeq}{\end{equation}}
\newcommand{\beqa}{\begin{eqnarray}}
\newcommand{\eeqa}{\end{eqnarray}}

\renewcommand\Re{\operatorname{Re}}

\usepackage{soul}
\usepackage[normalem]{ulem}

\graphicspath{{images/}}

\begin{document}

\title{An impurity in a heteronuclear two-component Bose mixture}
\author{G. Bighin$^{1,2}$, A. Burchianti$^{3,4}$, F. Minardi$^{3,4,5}$, and T. Macr\`{i}$^{6}$}
\affiliation{
$^{1}$IST Austria (Institute of Science and Technology Austria), Am Campus 1, 3400 Klosterneuburg, Austria\\
$^{2}$Institut f\"ur Theoretische Physik, Universit\"at Heidelberg, Philosophenweg 19, D-69120 Heidelberg, Germany\\
$^3$CNR-INO, Istituto Nazionale di Ottica, 50019 Sesto Fiorentino, Italy\\
$^4$European Laboratory for Nonlinear Spectroscopy - LENS, 50019 Sesto Fiorentino, Italy\\ 
$^{5}$Dipartimento di Fisica e Astronomia, Universit\`a di Bologna, 40127 Bologna, Italy\\
$^{6}$Departamento de F\'{i}sica Te\'{o}rica e Experimental, and International Institute of Physics,  Universidade Federal do Rio Grande do Norte, Campus Universit\'{a}rio, Lagoa Nova, Natal-RN 59078-970, Brazil
}

\date{\today}


\begin{abstract}
We study the fate of an impurity in an ultracold heteronuclear Bose mixture, focusing on the experimentally relevant case of a $^{41}$K-$^{87}$Rb mixture, with the impurity in a $^{41}$K hyperfine state. Our work provides a comprehensive description of an impurity in a BEC mixture with contact interactions across its phase diagram. We present results for the miscible and immiscible regimes, as well as for the impurity in a self-bound quantum droplet. Here, varying the interactions, we find novel, exotic states where the impurity localizes either at the center or at the surface of the droplet.
\end{abstract}

\maketitle

{\it Introduction. --}
The problem of a mobile impurity hosted in -- and interacting with -- a reservoir is a paradigm of many-body quantum theory \cite{Devreese:1996}. In general, interactions alter the properties of the impurity, starting from its inertia, in a way that critically depends on the excitation spectrum of the reservoir. Early on, this problem appeared when considering a single electron immersed in the environment of the ion lattice vibrations \cite{Landau:1933,Landau:1948,Frohlich:1954ds,Feynman:1955zz}, now known as a polaron, more specifically a ``Bose polaron'' to indicate that the environment is composed of bosonic modes, the lattice phonons \cite{Grusdt:2014}, and recently observed in \cite{Hu:2016in,Jorgensen:2016jr}.
In the last few years, initial studies have addressed the Bose polaron in a host system composed by two bosonic species, i.e.~a Bose mixture \cite{Compagno:2017, Ashida:2018ig,Boudjemaa:2020,Abdullaev:2020,Keiler:2021}, whose spectrum is much richer than its single-species counterpart. A remarkable property of Bose mixtures is the possibility to
form liquid-like self-bound droplets, arising from the interplay of mean-field attraction and beyond-mean-field repulsion \cite{Petrov:2015kh,Petrov:2016id}. Tuning mean-field interactions through a Feshbach resonance, quantum droplets have been observed in a homonuclear spin mixture of $^{39}$K, both in the presence of an external potential \cite{Cabrera:2018jn, Cheiney2018} and in free space \cite{Semeghini:2018bi}, as well as in  heteronuclear mixtures of $^{41}$K-$^{87}$Rb \cite{DErrico:2019wy} and $^{23}$Na-$^{87}$Rb \cite{Guo:2021}. Quantum droplets, arising from the competition between contact and long-range interactions \cite{reviewTommaso}, have been also observed in magnetic gases \cite{kadau2016,FerrierBarbut2016,Schmitt2016,Ferlaino2016,FerrierBarbutJPB2016,Wenzel2017} and, recently studied for dipolar mixtures
\cite{smith2021quantum,bisset2021quantum,lee2021miscibility}.

In this work we provide the first comprehensive description of a mobile impurity in a (heteronuclear) Bose mixture of atoms with contact interactions: we calculate the phase diagram of the impurity in a realistic case where, in proximity of a Feshbach resonance, an external magnetic field controls the interaction strength between the components of the Bose mixture \cite{Thalhammer:2008do}. 
First, we apply a generalized variational ansatz to compute the impurity spectral function and the impurity energy in the miscible and immiscible regimes. 
Then, we explicitly derive the beyond mean-field correction to the impurity-mixture interaction, leading to an effective interaction potential with variable sign and supporting several surface bound states.
Our findings provide access to important information for the study and the detection of Bose polarons in collisionally stable and long-lived Bose mixtures with far-reaching implications for future research.

\begin{figure}[!tbp]
\centering
    \includegraphics[width=.995\linewidth]{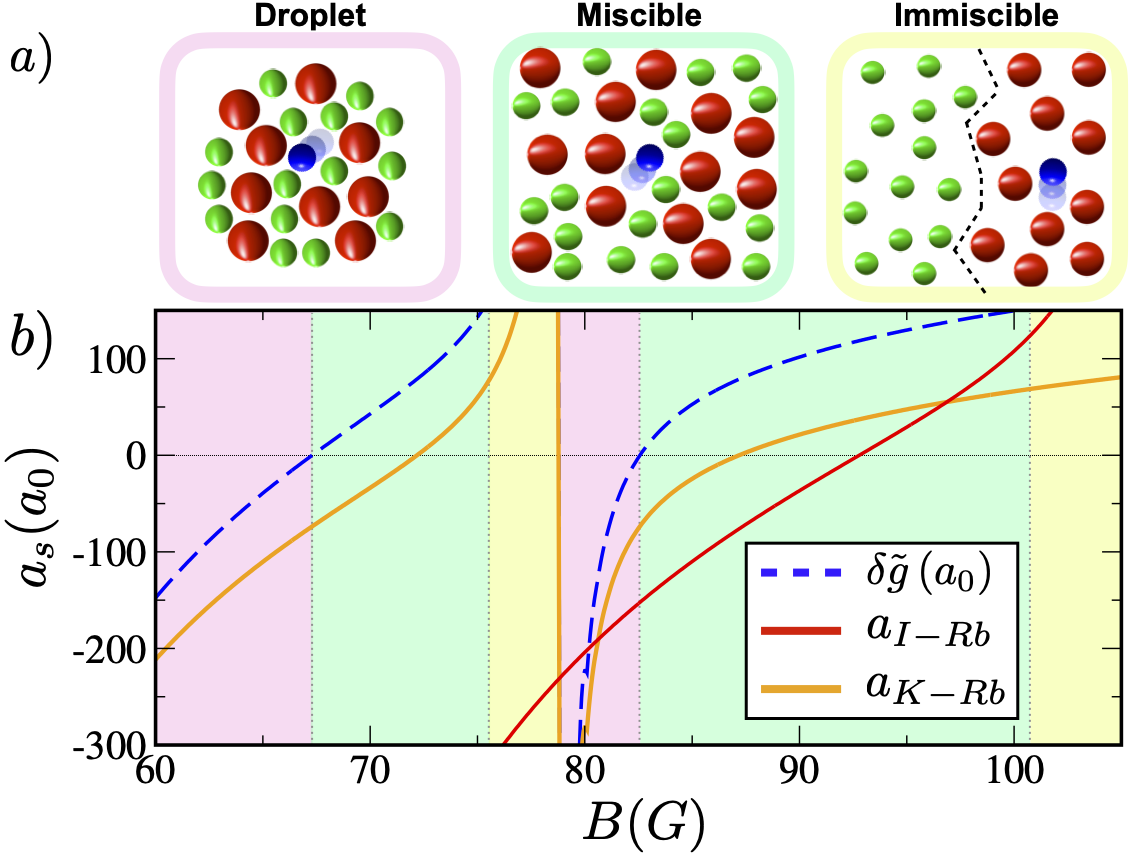}
\caption{{\bf Mobile impurity in a heteronuclear $^{87}$Rb-$^{41}$K mixture.}
(a) Quantum phases of the mixture: self-bound droplet, miscible phase, immiscible phase. In the self-bound phase (pink region) the mixture forms a droplet, stabilized by quantum fluctuations, on which the impurity can reside, see main text. In the miscible phase (green region) the impurity lives in a bosonic mixture. Depending on the interaction strength between the $^{41}$K impurity and each one of the mixture components in the immiscible, phase-separated regime (yellow region) the impurity occupies either one of the two domains. (b) Tunable scattering lengths $a_{\text{I-Rb}}$ between the impurity and $^{87}$Rb (red) and the inter-component scattering length $a_{\text{K-Rb}}$ (orange) across the magnetic field interval $B \in [60,105]$ G,  
note the Feshbach resonance at $B=78.9$ G.
In this magnetic field range, $a_{\text{I-K}}$, $a_{\text{Rb-Rb}}$, and $a_{\text{K-K}}$ are approximately constant (see main text).
(Dashed blue) Effective scaled mean-field coupling $\delta \tilde g=\delta g\, \sqrt{m_1\,m_2}/4\pi \hbar^2$ of the two-component mixture. When $\delta g<0$ the system is unstable toward collapse and it is stabilized by quantum fluctuations into a self-bound droplet phase (pink region).}\label{fig:one}
\end{figure}

\begin{figure}[!tbp]
\centering
    \includegraphics[width=.995\linewidth]{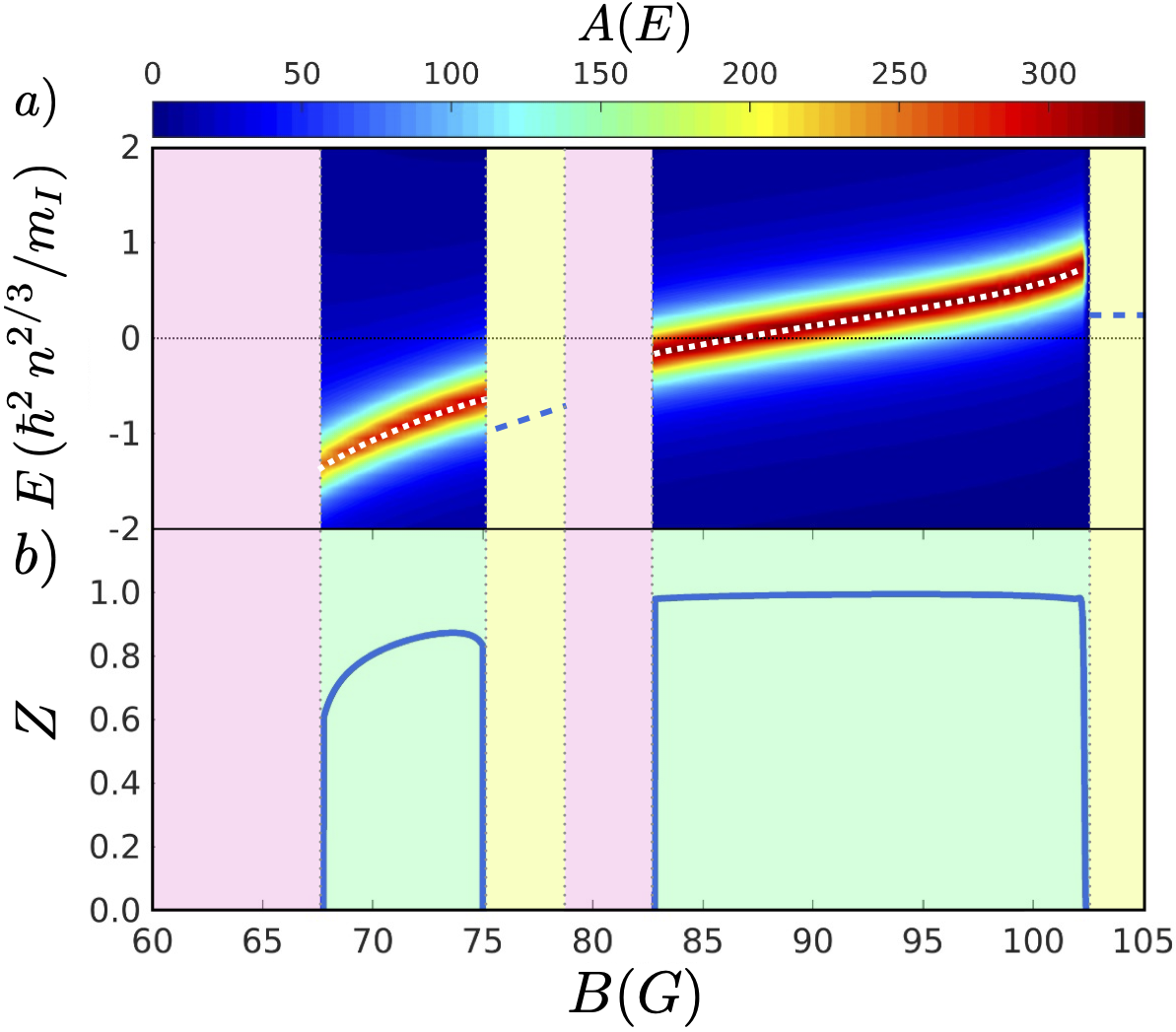}
\caption{{\bf Impurity properties in the miscible and immiscible phase.} (a) Polaron spectral
function $A(E)$ in the miscible phase of the mixture, as a function of the external magnetic field B and of the polaron energy. The dotted white line shows the mean-field solution of the equations of motion, whereas the dashed blue line in the immiscible region shows the energy of the polaron coupled to a single component, see main text. Energies are scaled by the peak density $n$ of the mixture. (b) Quasiparticle weight $Z$ in the immiscible region, as a function of the magnetic field B.
}
\label{fig:two}
\end{figure}

{\textit{The system. --}} Let us consider a two-component, ultracold, interacting Bose-Bose mixture \cite{Larsen:1963cj}. The strength of the interspecies contact interaction is determined by the parameter $g_{12}=2 \pi \hbar^2 a_{12} / \mu$ where $a_{12}$ is the interspecies scattering length, $\mu=m_1 m_2 / (m_1 + m_2)$ is the reduced mass, $m_i$ is the mass of $i$-th species bosons. The strength of the intraspecies contact interaction is determined by $g_{ii}=4 \pi \hbar^2 a_{ii} / m_i$, $i=1,2$, with $a_{ii}$ the intraspecies scattering length for the $i$-th species, so that the system can be described by the Hamiltonian
\begin{multline}
\hat{H}_\text{bb} = \int \mathrm{d}^3 r \ \sum_{i=1,2} \hat{\phi}^\dagger_i (\mathbf{r}) \left(-\frac{\hbar^2 \nabla^2}{2 m_i} + \frac{g_{ii}}{2} |\hat{\phi}_i (\mathbf{r})|^2\right) \hat{\phi}_i (\mathbf{r}) + \\
+ g_{12} \int \mathrm{d}^3 r |\hat\phi_1 (\mathbf{r})|^2|\hat\phi_2 (\mathbf{r})|^2
\end{multline}
where the $\hat{\phi}_i^\dagger (\mathbf{r})$ ($\hat{\phi}_i (\mathbf{r})$) fields operators create (annihilate) a bosonic field excitation at position $\mathbf{r}$ in the $i$-th component, respectively. In addition to this, we also consider a third component in the impurity limit, i.e.~a third component much more dilute than the other two, so that we can neglect inter-component interactions and we can describe it in the first quantization formalism via operators describing the impurity position $\mathbf{\hat{R}}$ and momentum $\mathbf{\hat{P}}$. The Hamiltonian describing the impurity motion and the interaction with the other two components reads
\begin{equation}
\hat{H}_\text{I} = \frac{\hat{\mathbf{P}^2}}{2m_\text{I}} + \sum_{i=1,2} g_{Ii} \int \mathrm{d}^3 r \ \rho(\mathbf{r}) \ |\hat{\phi}_i (\mathbf{r})|^2
\end{equation}
where $\rho(\mathbf{r}) = \delta^{(3)} (\mathbf{r} - \mathbf{\hat{R}})$, $m_I$ is the impurity mass, $g_{Ii}=2 \pi \hbar^2 a_{Ii} / \mu_{Ii}$ where $a_{Ii}$ is the scattering length between the impurity and the $i$-th component, and $\mu_{Ii}=m_I m_i / (m_I + m_i)$.

{\textit{Impurity in the miscible and immiscible phases. --}} In order to describe the miscible phase we begin by expanding the field operators  in the plane wave basis $\hat \phi_1 (\mathbf{r}) = V^{-1/2} \sum_\mathbf{q} e^{i \mathbf{q}\cdot \mathbf{r}} \alpha_\mathbf{q}$, $\hat \phi_2 (\mathbf{r}) = V^{-1/2} \sum_\mathbf{q} e^{i \mathbf{q}\cdot \mathbf{r}} \beta_\mathbf{q}$. We subsequently employ Bogoliubov approximation, considering a macroscopic occupation of the ground state for each species and linear fluctuations around it, so that the Hamiltonian can be brought to a diagonal form by means of a generalized $4 \times 4$ Bogoliubov transformation of the $\alpha_\mathbf{q}$, $\beta_\mathbf{q}$ fields \cite{Larsen:1963cj}, at the expense of switching to a new basis where the two components are mixed as to form two new, effective components that we shall dub A and B. 
In this new basis, the diagonal Hamiltonian reads
\beq
H_\text{bos} = \sum_{\mathbf{k}} \hbar \omega^{(A)}_\mathbf{k} \hat{a}^\dagger_{\mathbf{k}} \hat{a}_{\mathbf{k}} + \sum_{\mathbf{k}} \hbar \omega^{(B)}_\mathbf{k} \hat{b}^\dagger_{\mathbf{k}} \hat{b}_{\mathbf{k}}
\eeq
where $\omega^{(i)}_{\mathbf{k}}$ is the effective dispersion relation for the $i$t-th component and the $\hat{a}^\dagger_\mathbf{k}$ ($\hat{b}^\dagger_\mathbf{k}$) operator creates a Bogoliubov excitation for the A (B) component, respectively.

The mobile impurity in the third component is described by $\hat{H}_\text{imp} = \hat{\mathbf{P}}^2 / 2 m_I$ and the impurity couplings to the A and B components of the bosonic bath read at the linear level
\begin{equation}
\hat{H}^{(1)}_\text{imp-bos} = \sum_{\mathbf{k} \neq 0} e^{\mathrm{i} \mathbf{k} \cdot \hat{\mathbf{R}}} [ U_A(\mathbf{k}) (\hat{a}_\mathbf{k} + \hat{a}^\dagger_{-\mathbf{k}}) + U_B(\mathbf{k}) (\hat{b}_\mathbf{k} + \hat{b}^\dagger_{-\mathbf{k}}) ] \; ,
\label{eq:h1}
\end{equation}
the effective potentials $U_A(\mathbf{k})$ and $U_B(\mathbf{k})$ being a linear superposition of the microscopic impurity-bath potentials, depending also on the interaction parameters $g_{ij}$, as derived in \cite{SM}.
It has been shown \cite{Rath:2013hu,Shchadilova:2016jz} that terms bilinear in the bosonic operators, describing the scattering of the impurity off the condensate, are important for an accurate description of impurities in ultracold gases. For this reason, we include bilinear terms
\beq
H^{(2)}_\text{imp-bos} = \sum_{\substack{\mathbf{k}, \mathbf{k}'\\i=A,B}} e^{\mathrm{i} (\mathbf{k}+\mathbf{k}') \cdot \hat{\mathbf{R}}} \ \Psi_a (\mathbf{k'}) \mathbb{M}^{i}_{a b} (\mathbf{k}',\mathbf{k}) \Psi_b (\mathbf{k}),
\label{eq:h2}
\eeq
having grouped the creation and annihilation operators into a spinor-like object $\Psi (\mathbf{k})=(a_\mathbf{k} \ a^\dagger_{-\mathbf{k}} \ b_\mathbf{k} \ b^\dagger_{-\mathbf{k}} )^T$ and the mixing matrices $\mathbb{M}^{i}_{a b} (\mathbf{k}',\mathbf{k})$ \cite{SM}.

We now want to solve the full Hamiltonian $\hat{H} = H_\text{bos} + H_\text{imp} + \hat{H}^{(1)}_\text{imp-bos} + H^{(2)}_\text{imp-bos}$. In order to do so, at first, we make use of a canonical transformation $\hat{S}=\exp \small( \mathrm{i} \hat{\mathbf{R}} \cdot \hat{\mathbf{P}}_A/\hbar \small) \exp \small( \mathrm{i} \hat{\mathbf{R}} \cdot \hat{\mathbf{P}}_B /\hbar \small)$ so that the transformed Hamiltonian $\mathcal{H} = \hat{S}^{-1} \hat{H} \hat{S}$ describes the system in a frame of reference comoving with the impurity. Here $\hat{\mathbf{P}}_A = \sum_{\mathbf{k}} \ \hbar\mathbf{k} \ \hat{a}^\dagger_{\mathbf{k}} \hat{a}_{\mathbf{k}}$ and the similarly-defined $\hat{\mathbf{P}}_B$ are the bosonic momenta in the $A$ and $B$ component, respectively, that we use as generators of spatial translations for bosons. We can now study the dynamics of the system by means of a time-dependent variational ansatz \cite{Kramer:1981,Shchadilova:2016jz,Ashida:2018ig,Ardila:2021}
\beq
\ket{\Psi (t)} = e^{i \phi(t)} e^{\sum_{\mathbf{k}} \alpha_\mathbf{k} (t) a^\dagger_\mathbf{k} + \beta_\mathbf{k} (t) b^\dagger_\mathbf{k} - h.c.} \ket{0}_\text{bos}^{A} \ket{0}_\text{bos}^{B}
\label{eq:ansatz}
\eeq
where $\ket{0}_\text{bos}^{i}$ is the boson vacuum for the $i$ component. The coherent-state ansatz of Eq. (\ref{eq:ansatz}) constitutes an exact solution for the ground state of an infinite-mass impurity. We subsequently numerically determine the variational coefficients $\alpha_\mathbf{k} (t)$, $\beta_\mathbf{k} (t)$ via the Euler-Lagrange equation obtained from the Lagrangian $\mathscr{L} = \matrixel{\Psi (t)}{\mathrm{i} \hbar\partial_t - \hat{\mathcal{H}}}{\Psi (t)}$. The time evolution of the time-dependent phase $\phi(t)$, on the other hand, is found by projecting Schr\"odinger equation onto the chosen variational wave function, i.e.~by evaluating $\matrixel{\Psi (t)}{\mathrm{i} \hbar \partial_t}{\Psi (t)} = \matrixel{\Psi (t)}{\hat{\mathcal{H}}}{\Psi (t)}$ and numerically solving for $\phi(t)$. Finally, the dynamical overlap or Loschmidt echo $S(t) = \matrixel{\Psi(0)}{e^{-\mathrm{i} \hat{\mathcal{H}} t/\hbar}}{\Psi(0)} = \braket{\Psi(0)}{\Psi(t)}$ contains full information about the spectrum of the system, allowing one to immediately obtain the spectral function as $A(\omega) = 2 \Re \int_0^\infty \mathrm{d} t \ e^{\mathrm{i} \omega t} S(t)$.

As a concrete realization of this system, motivated by recent experiments \cite{Thalhammer:2009co, Burchianti:2020du, DErrico:2019wy}, we consider a heteronuclear $^{41}$K-$^{87}$Rb Bose mixture -- which we shall dub species `$1$' and `$2$', respectively -- on top of which we consider a dilute third component realized with a different hyperfine state of $^{41}$K -- dubbed `I' species. The atoms forming the bosonic reservoir are in their hyperfine ground state, $(F=1, m_F=1)$ for both species, while the $^{41}$K impurity is in the second-lowest hyperfine state $(F=1, m_F=0)$; this specific configuration is not affected by spin-exchange collisions, which generally restrict the lifetime of atomic mixtures. In the impurity limit for the third component, the system is described by five scattering lengths, namely $a_\text{K-K}$, $a_\text{K-Rb}$, $a_\text{Rb-Rb}$, $a_\text{I-K}$, $a_\text{I-Rb}$. Importantly, these scattering lengths are all known, and $a_\text{K-Rb}$ and $a_\text{I-Rb}$ can be tuned thanks to experimentally accessible Feshbach resonances \cite{Ferlaino:2006fe,DErrico:2007fe, Simoni:2008ne, Thalhammer:2009co, Thalhammer:2008do}: in Fig.~\ref{fig:one} we display the behaviour of $a_\text{K-Rb}$ and $a_\text{I-Rb}$ as a function of the magnetic field $B$ in the range $B\in [60, 105]$ G. The other three scattering lengths are almost constant in the range considered, i.e.~$a_\text{K-K} \simeq a_\text{I-K}  \simeq 62\, a_0$, $a_\text{Rb-Rb}  \simeq 100.4\, a_0$.

The liquid-gas transition parameter $\delta g = g_{12} + \sqrt{g_{11} g_{22}}$, also shown in Fig.~\ref{fig:one}, 
allows us to chart 
the Bose mixture phase diagram:
as the magnetic field is varied in the aforementioned range, the mixture goes through the droplet, miscible and immiscible phases. The main result of the present Letter 
is the analysis the fate of the impurity across this 
phase diagram.

In Fig.~\ref{fig:two} we consider this heteronuclear 
Bose mixture, 
and we plot the impurity spectral function $A(E)$ as a function of the (scaled) energy $E$ and of the magnetic field $B$ in the miscible phase.
In the whole 
range of magnetic field, 
both polaron couplings are off-resonant, so that the many-body environment simply shifts the energy of a sharp quasiparticle peak -- Fig.~\ref{fig:two}a -- while maintaining a relatively large quasiparticle weight $Z$ -- Fig.~\ref{fig:two}b. 
We note that the energy of the quasiparticle peak is well approximated by the mean-field solution (dotted white line) obtained by setting $\dot{\alpha}_\mathbf{k} (t) = \dot{\beta}_\mathbf{k} (t) = 0$ in the equations of motion. In the same figure we also characterize the immiscible phase polaron energy (dashed blue) considering that the polaron will reside in the most energetically favorable component.

\begin{figure}[!tbp]
\centering
    \includegraphics[width=.995\linewidth]{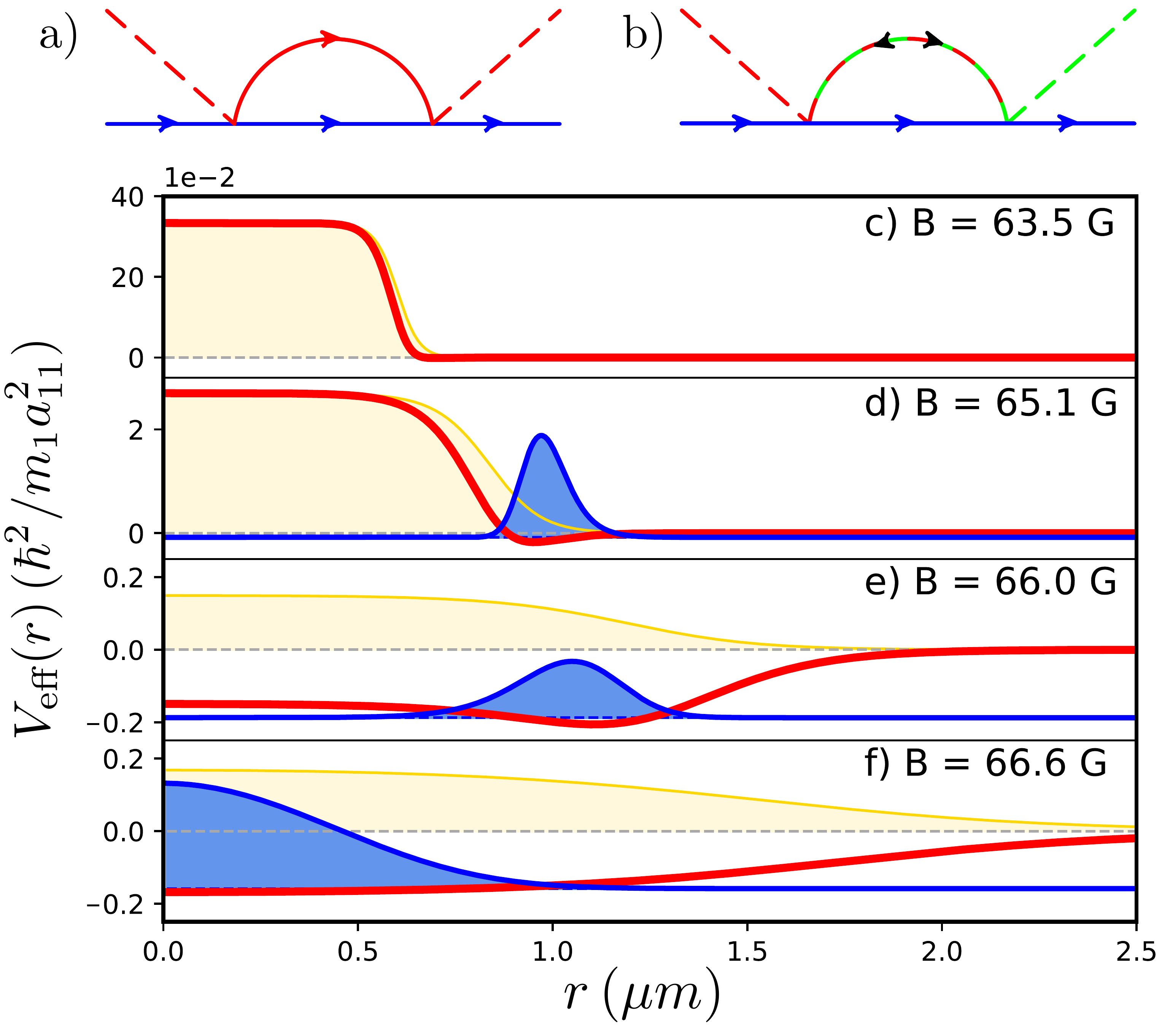}
\caption{{\bf Localized impurity in a self-bound droplet.}
a) and b) Representative impurity-droplet Feynman diagrams at second-order perturbation theory in the small parameters $a_{Ii}/\xi_i$, $i=A,B$. The red and green dashed lines refer to the first and second condensate component, respectively, while the blue lines represent impurity propagators. Note that diagrams such as the one in b) can `mix' the condensate components, by annihilating an excitation in the first BEC component and creating one in the second, or vice-versa. 
c)-f) Effective impurity potential $V_\text{eff}(r)$ (red curves) for an impurity in a self-bound droplet (yellow shaded region). The impurity density (blue shaded) and its ground-state energy (blue dashed lines) are shown for several magnetic fields.
c) $B=63.5$ G the potential does not support bound states in three dimensions. d) $B=65.1$ G and e) $B=66.0$ G the impurity is localized at the surface of the droplet at a distance $r \approx 1 \mu$m form the center. f) $B=66.6$ G the impurity is centered in the self-bound droplet.}
\label{fig:three}
\end{figure}

\begin{figure*}[t]
\centering
    \includegraphics[width=.995\linewidth]{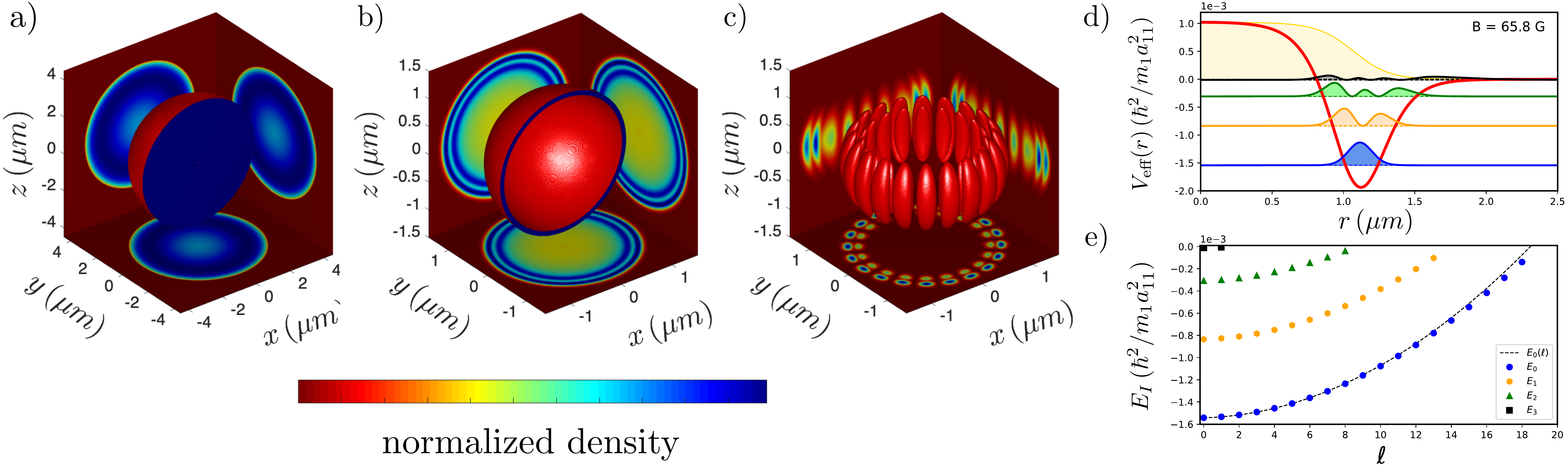}
\caption{{\bf Densities and  excited states of the impurity.}
a-c) Impurity density of the ground and excited states. Main figures: isosurfaces of constant density. Planes display integrated densities along the axis orthogonal to the plane. Densities (colorbar) are scaled by the maximum density of each configuration. The number of particles in the droplet supporting these bound states is $N_D=4\times 10^4$.
a) Ground state of an impurity centered in the droplet at $B=66.6$ G.
b) Ground state of an impurity localized at the droplet surface at $B=65.8$ G.
c) Excited state of an impurity at $B=65.8$ G for $\ell=10$ and $m=10$. 
d) Effective potential $V_\text{eff}(r)$ and density of the impurity $n_I(r)$ for the $n=0,\dots,3$ s-wave bound states for a spherical droplet with density $n_D(r)$ (yellow). 
The eigenstates are localized on the surface of the droplet at $r\approx\, 1.1 \mu$m from the droplet center.
The density of the droplet is scaled by a constant factor for illustrative purposes. The ground state (blue) is also shown in b)
e) Spectrum of the impurity eigenstates in the presence of the effective potential $V_\text{eff}$(r) of a).
The black dashed line is the analytical prediction $E_I(\ell)-E_I(0)\propto \ell(\ell+1)$ discussed in the text.
Eigenstates with $n=1$ and $\ell=0$ and $\ell=10$, $m=10$ are shown in b) and c) respectively.
The parameters for d) and e) are $N_D=4\times 10^4$ and $B=65.8$ G.}
\label{fig:four}
\end{figure*}

{\textit{Impurity in a self-bound droplet. --}} We now draw our attention to the droplet phase. We describe the self-bound droplet within the Gross-Pitaevskii (GP) formalism, the two BEC components being described by complex fields $\phi_i(\mathbf{r})$ with the associated energy functional
\begin{multline}
E_{bb} [\phi_i] = \int \mathrm{d}^3 r 
\sum_{i=1,2} \left(
\frac{\hbar^2|\nabla \phi_i|^2}{2m_i} 
+\frac{g_{ii}}{2}|\phi_i|^4
\right)+ \\
+ g_{12}|\phi_1|^2|\phi_2|^2 +
\frac{8}{15\pi^2\hbar^3}
\left(
m_1^\frac{3}{5}g_{11}|\phi_1|^2+
m_2^\frac{3}{5}g_{22}|\phi_2|^2
\right)^\frac{5}{2} \; .
\end{multline}
where the last term is the beyond mean-field interaction for a general two-component mixture \cite{Minardi:2019ef,PhysRevLett.126.115301}. The impurity and the interaction between the impurity and the Bose mixture are described by the energy functional
\begin{equation}
\label{Eimp}
E_{I}= \int \mathrm{d}^3 r
\frac{\hbar^2|\nabla \psi|^2}{2m_I}
+ 
\left(g_{ID} |\phi (\mathbf{r})|^2 + \mathscr{E}_\text{BMF} (\mathbf{r})\right) 
|\psi (\mathbf{r})|^2,
\end{equation}
where now the impurity component is described by the wavefunction $\psi (\mathbf{r})$, the total density 
$n(\mathbf{r})\equiv |\phi(\mathbf{r})|^2=n_1(\mathbf{r}) + n_2(\mathbf{r})$, and $g_{ID}$ is an effective mean-field coupling constant \cite{SM}. 
The last term
$\mathscr{E}_\text{BMF} (\mathbf{r})$ is the beyond mean-field interaction for a general two-component mixture. We obtain it by means of perturbation theory in the small parameters $(a_{Ii}/\xi_i)$, $i=1,2$, where $\xi_i = 1 / \sqrt{8 \pi n_i a_{ii}}$ is the healing length for the $i$-th component and 
$n_i$ is its density. 
This approach is equivalent to summing $16$ different Feynman diagrams corresponding to $2$nd-order processes; some of the diagrams do not mix the condensate components -- as in Fig.~\ref{fig:three}a \cite{Christensen2015quasiparticle}, whereas our binary mixture also allows for processes mixing the two components, as exemplified in Fig.~\ref{fig:three}b. 
We 
report here only the final results \cite{SM}. Within the local-density approximation, one obtains

\begin{multline}
\label{Ebmf}
\mathscr{E}_\text{BMF}(\mathbf{r}) = \left( \frac{2 \pi \hbar^2 \xi_1 n}{\mu_{I1}} \frac{1}{1+\alpha} \right) \left( \frac{a_{I1}}{\xi_1} \right)^2 \frac{m_1}{\mu_{I1}} I_1 + \\
+ \left( \frac{2 \pi \hbar^2 \xi_2 n}{\mu_{I2}} \frac{\alpha}{1+\alpha} \right) \left( \frac{a_{I2}}{\xi_2} \right)^2 \frac{m_2}{\mu_{I2}} I_2
\end{multline}
where the total density $n(\mathbf{r})$ and the healing lengths $\xi_i(\mathbf{r})$ 
are evaluated at the impurity position and $I_i,\, i=1,2$ are dimensionless regularized integrals depending on the condensate Bogoliubov amplitudes, and the coupling constant ratio $\alpha=\sqrt{g_{11}/g_{22}}$.

To probe the equilibrium properties of the impurity in a droplet environment, we set the relative number of particles in each of the two components to satisfy the constraint
$N_1 / N_2 = \sqrt{g_{22}/g_{11}}$ \cite{Petrov:2015kh}. Rescaling lengths by $a_{11}$ and energies by $E_1 = \hbar^2 / (m_1 a_{11}^2)$, we 
write a set of coupled generalized GP equations for the impurity-droplet system and we 
again analyze the concrete case of the heteronuclear $^{41}\mathrm{K}$ - $^{87}\mathrm{Rb}$ mixture introduced above. In Fig.~\ref{fig:three} c)-f) we report the radial density profiles for the condensate (yellow shaded region) and for the impurity (blue region) for four different values of the magnetic field. We also plot the effective potential exerted by the mixture on the impurity, $V_\text{eff} (\mathbf{r}) = g_{ID}|\phi (\mathbf{r})|^2 + \mathscr{E}_\text{BMF} (\mathbf{r})$, sum of the mean-field term linear in the droplet density $n(\mathbf{r})$, and $\mathscr{E}_\text{BMF} (\mathbf{r})$ scaling as $n(\mathbf{r})^{3/2}$. The latter is repulsive, while the former can be attractive when $g_{ID}<0$: in this case, occurring for our specific mixture, $V_\text{eff}(\mathbf{r})$ is repulsive (attractive) in the high (low) density region of the droplet, respectively, giving rise to a rich phenomenology.
For $B = 63.5$ G, in three dimensions $V_\text{eff}(\mathbf{r})$ does not support bound states. On the other hand, as the magnetic field is increased, for $B = 65.1$ G and for $B = 66.0$ G we observe that the impurity is localized at the surface of the droplet 
(Fig.~\ref{fig:four}b). 
Finally, as the magnetic field is further increased, we show that for $B = 66.6$ G the impurity is localized at the center of the self-bound droplet 
Interestingly, these surface and center bound states occur in a range of magnetic fields where long-lived droplets have been already produced \cite{DErrico:2019wy, Burchianti:2020du}.  In current experiments the existence of such states, in which the impurity either localizes at the center of the droplet or at its surface, could be probed by performing high-resolution imaging \cite{Guo:2021}. 
Although we deal with the case of a single impurity, we expect that the results are not substantially affected for a small -- but detectable -- 
number of impurities, e.g. a few percent of the droplet atom number.

Finally, in Fig.~\ref{fig:four} c)-e) we study excited states of the impurity.
Depending on the magnetic field strengths few bound states appear, their number depending also on the impurity angular momentum $\ell$. 
We notice that, for localized states at the surface, the lowest-energy impurity states of finite $\ell$ is well described by $E_I(\ell) = E_I(0)+ \hbar^2\ell(\ell+1)/2m_I r_0^2$ (black dashed line in Fig.~(\ref{fig:four}e), where $r_0$ is the droplet radius: 
the centrifugal barrier affects localized surface states at finite $\ell$ providing essentially a constant energy shift with respect to the $\ell=0$ case.

{\it Conclusions and outlook. -- } In this work we studied the effect of an impurity in a two-component heteronuclear Bose mixture. We found exotic surface states both for the ground and the excited states of the impurity, and characterized the miscible phase via the impurity spectral function, which is readily accessible in current experiments.
Our findings provide access to relevant information for the study and the detection of Bose polarons in collisionally stable and long-lived Bose mixtures such as $^{41}$K-$^{87}$Rb. This study has far-reaching implications for further research, 
e.g.~by considering a similar scenario with fermionic impurities \cite{Wenzel:2018cr}, a finite Rabi coupling between the two BEC components \cite{Cappellaro:2017ia}, lower dimensionalities \cite{Mistakidis:2021}, the coupling to highly-excited Rydberg states \cite{Schmidt:2016ge}, or heliophobic impurities residing on the surface of a $^4$He nanodroplet \cite{Toennies:2014}.

\begin{acknowledgments}
\textit{Acknowledgments.}
We thank A.~Simoni for providing the calculations of the  intercomponents scattering lengths. We gratefully acknowledge stimulating discussions with L.A.~Pe\~{n}a Ardila, R.~Schmidt, H.~Silva, V.~Zampronio, and M.~Prevedelli for careful reading. 
G.B.~ acknowledges support from the Austrian Science Fund (FWF), under project No.~M2641-N27.
T.M.~ acknowledges CNPq for support through 
Bolsa de produtividade em Pesquisa n.311079/2015-6. 
This work is supported by the Deutsche Forschungsgemeinschaft (DFG, German Research Foundation) under Germany's Excellence Strategy EXC2181/1-390900948 (the Heidelberg STRUCTURES Excellence Cluster).
This work was supported by the Serrapilheira Institute 
(grant number Serra-1812-27802).
We thank the High Performance Computing Center (NPAD) at UFRN for providing computational resources.
\end{acknowledgments}

\clearpage
\pagebreak
\widetext
\begin{center}
\textbf{\large Supplemental Material:\\[5pt] An impurity in a heteronuclear two-component Bose mixture}
\end{center}
\setcounter{equation}{0}
\setcounter{figure}{0}
\setcounter{table}{0}
\setcounter{page}{1}
\makeatletter
\renewcommand{\theequation}{S\arabic{equation}}
\renewcommand{\thefigure}{S\arabic{figure}}
\renewcommand{\bibnumfmt}[1]{[S#1]}
\renewcommand{\citenumfont}[1]{S#1}
\makeatother

\section{From the density-density interaction to the effective Hamiltonian}

The density-density interaction term describing a first-quantized impurity in a bosonic many-body bath interacting with two different bosonic species is given by
\beq
\hat{H}_\text{imp-bos} = \sum_{\mathbf{k}, \mathbf{q}} V_A(\mathbf{q}) \hat{\rho} (\mathbf{q}) \hat{\alpha}^\dagger_{\mathbf{k} - \mathbf{q}} \hat{\alpha}_\mathbf{k} + \sum_{\mathbf{k}, \mathbf{q}} V_B(\mathbf{q}) \hat{\rho} (\mathbf{q}) \hat{\beta}^\dagger_{\mathbf{k} - \mathbf{q}} \hat{\beta}_\mathbf{k}
\label{eq:densitydensity}
\eeq
where $V(\mathbf{q})$ is the Fourier transform of the impurity-bath potential, the $\alpha^\dagger_\mathbf{k}$ ($\beta^\dagger_\mathbf{k}$) creates a bosonic excitation in the first (second) component respectively. Finally $\hat{\rho}(\mathbf{q}) = \exp \small( \mathrm{i} \mathbf{q} \cdot \hat{\mathbf{R}} \small)$ is the Fourier transform of the density of an impurity located at position $\hat{\mathbf{R}}$. Bogoliubov approximation consists in separating the macroscopic occupation of the ground state from the fluctuations
\beq
\hat{\alpha}_\mathbf{k} = (2 \pi)^3 \sqrt{n_A} \delta(\mathbf{k}) + \hat{A}_{\mathbf{k} \neq 0}
\label{eq:a0}
\eeq
subsequently retaining only terms linear in $\hat{A}$, neglecting higher order terms. A completely analogous procedure is employed for the $\hat{\beta}_\mathbf{k}$ operators. Subsequently, the Bogoliubov transformation brings the Hamiltonian in a diagonal form. In the present case the Bogoliubov transformation is given by a $4 \times 4$ matrix that rotates the creation and annihilation operators for the two species in the following way \cite{Larsen:1963cjSM}
\beq
\begin{pmatrix}
\hat{A}_\mathbf{k} \\
\hat{A}^\dagger_{-\mathbf{k}} \\
\hat{B}_\mathbf{k} \\
\hat{B}^\dagger_{-\mathbf{k}} \\
\end{pmatrix} =
\begin{pmatrix}
M^{11}_\mathbf{k} & M^{12}_\mathbf{k} & M^{13}_\mathbf{k} & M^{14}_\mathbf{k} \\
M^{21}_\mathbf{k} & M^{22}_\mathbf{k} & M^{23}_\mathbf{k} & M^{24}_\mathbf{k} \\
M^{31}_\mathbf{k} & M^{32}_\mathbf{k} & M^{33}_\mathbf{k} & M^{34}_\mathbf{k} \\
M^{41}_\mathbf{k} & M^{42}_\mathbf{k} & M^{43}_\mathbf{k} & M^{44}_\mathbf{k}
\end{pmatrix} \begin{pmatrix}
\hat{a}_\mathbf{k} \\
\hat{a}^\dagger_{-\mathbf{k}} \\
\hat{b}_\mathbf{k} \\
\hat{b}^\dagger_{-\mathbf{k}} \\
\end{pmatrix} \; .
\label{eq:brotation}
\eeq
A full derivation of the coefficients of the Bogoliubov transformation for a general heteronuclear mixture considered in the text is a lengthy calculation. We refer the reader to \cite{Larsen:1963cjSM} for the details.
In the case of Eq.~(\ref{eq:densitydensity}) the separation of Eq.~(\ref{eq:a0}) gives
\beq
\hat{H}^{(1)}_\text{imp-bos} = \sqrt{n_A} \sum_{\mathbf{k} \neq 0} V_A(\mathbf{k}) e^{\mathrm{i} \mathbf{k} \cdot \hat{\mathbf{R}}} (\hat{A}_\mathbf{k} + \hat{A}^\dagger_{-\mathbf{k}}) + \sqrt{n_B} \sum_{\mathbf{k} \neq 0} V_B(\mathbf{k}) e^{\mathrm{i} \mathbf{k} \cdot \hat{\mathbf{R}}} (\hat{B}_\mathbf{k} + \hat{B}^\dagger_{-\mathbf{k}})
\eeq
having neglected higher order terms in $\hat{A}$ and $\hat{B}$ and having omitted a constant factor $n_A V_A(\mathbf{k}=0) + n_B V_B(\mathbf{k}=0)$. Bogoliubov transformation then gives

\beq
\begin{aligned}
\hat{H}^{(1)}_\text{imp-bos} &= \sqrt{n_A} \sum_{\mathbf{k} \neq 0} V_A(\mathbf{k}) e^{\mathrm{i} \mathbf{k} \cdot \hat{\mathbf{R}}} [ (M^{11}_\mathbf{k} + M^{21}_\mathbf{k}) \hat{a}_\mathbf{k} + (M^{12}_\mathbf{k} + M^{22}_\mathbf{k}) \hat{a}^\dagger_{-\mathbf{k}} ] + \\
&+\sqrt{n_A} \sum_{\mathbf{k} \neq 0} V_A(\mathbf{k}) e^{\mathrm{i} \mathbf{k} \cdot \hat{\mathbf{R}}} [(M^{13}_\mathbf{k} + M^{23}_\mathbf{k}) \hat{b}_\mathbf{k} + (M^{14}_\mathbf{k} + M^{24}_\mathbf{k}) \hat{b}^\dagger_{-\mathbf{k}} ] + \\
&+ \sqrt{n_B} \sum_{\mathbf{k} \neq 0} V_B(\mathbf{k}) e^{\mathrm{i} \mathbf{k} \cdot \hat{\mathbf{R}}} [ (M^{31}_\mathbf{k} + M^{41}_\mathbf{k}) \hat{a}_\mathbf{k} + (M^{32}_\mathbf{k} + M^{42}_\mathbf{k}) \hat{a}^\dagger_{-\mathbf{k}} ] + \\
&+ \sqrt{n_B} \sum_{\mathbf{k} \neq 0} V_B(\mathbf{k}) e^{\mathrm{i} \mathbf{k} \cdot \hat{\mathbf{R}}} [(M^{33}_\mathbf{k} + M^{43}_\mathbf{k}) \hat{b}_\mathbf{k} + (M^{34}_\mathbf{k} + M^{44}_\mathbf{k}) \hat{b}^\dagger_{-\mathbf{k}} ]
\end{aligned}
\eeq
Here one immediately sees that the Hermiticity condition implies (assuming $V_i (\mathbf{k}) = V_i (-\mathbf{k})$ for $i=A,B$ and that the matrix elements $M^{ij}_\mathbf{k}$ are real)
\beq
\begin{aligned}
M^{11}_\mathbf{k} + M^{21}_\mathbf{k} = M^{12}_\mathbf{k} + M^{22}_\mathbf{k} \\
M^{13}_\mathbf{k} + M^{23}_\mathbf{k} = M^{14}_\mathbf{k} + M^{24}_\mathbf{k} \\
M^{31}_\mathbf{k} + M^{41}_\mathbf{k} = M^{32}_\mathbf{k} + M^{42}_\mathbf{k} \\
M^{33}_\mathbf{k} + M^{43}_\mathbf{k} = M^{34}_\mathbf{k} + M^{44}_\mathbf{k}
\end{aligned}
\eeq
Finally, one can rewrite the interaction term as
\beq
\hat{H}^{(1)}_\text{imp-bos} = \sum_{\mathbf{k} \neq 0} U_A(\mathbf{k}) e^{\mathrm{i} \mathbf{k} \cdot \hat{\mathbf{R}}} (\hat{a}_\mathbf{k} + \hat{a}^\dagger_{-\mathbf{k}}) + \sum_{\mathbf{k} \neq 0} U_B(\mathbf{k}) e^{\mathrm{i} \mathbf{k} \cdot \hat{\mathbf{R}}} (\hat{b}_\mathbf{k} + \hat{b}^\dagger_{-\mathbf{k}})
\eeq
having introduced the effective potentials we use in the main text
\begin{align}
U_A(\mathbf{k}) &= \sqrt{n_A} V_A(\mathbf{k}) (M^{11}_\mathbf{k} + M^{21}_\mathbf{k}) +  \sqrt{n_B} V_B(\mathbf{k}) (M^{31}_\mathbf{k} + M^{41}_\mathbf{k}) \; , \\
U_B(\mathbf{k}) &= \sqrt{n_B} V_B (\mathbf{k}) (M^{33}_\mathbf{k} + M^{43}_\mathbf{k}) + \sqrt{n_A} V_A (\mathbf{k}) (M^{13}_\mathbf{k} + M^{23}_\mathbf{k}) \;.
\end{align}
The total Fr\"ohlich-level Hamiltonian $\hat{H} = \hat{H}_\text{bos} + \hat{H}_\text{imp} + \hat{H}_\text{imp-bos} $ is then
\beq
\begin{split}
\hat{H} = & \frac{\hat{\mathbf{P}}^2}{2M} + \sum_{\mathbf{k}} \hbar\omega^{(A)}_\mathbf{k} a^\dagger_{\mathbf{k}} a_{\mathbf{k}} + \sum_{\mathbf{k}} \hbar\omega^{(B)}_\mathbf{k} b^\dagger_{\mathbf{k}} b_{\mathbf{k}} + \\ + &\sum_{\mathbf{k} \neq 0} U_A(\mathbf{k}) e^{\mathrm{i} \mathbf{k} \cdot \hat{\mathbf{R}}} (\hat{a}_\mathbf{k} + \hat{a}^\dagger_{-\mathbf{k}}) + \sum_{\mathbf{k} \neq 0} U_B(\mathbf{k}) e^{\mathrm{i} \mathbf{k} \cdot \hat{\mathbf{R}}} (\hat{b}_\mathbf{k} + \hat{b}^\dagger_{-\mathbf{k}})
\end{split}
\label{frolich}
\eeq

\section{Extended Hamiltonian}

We now include the higher-order terms, i.e. terms $\sim \hat{a}^\dagger \hat{a}^\dagger$ and similar couplings, describing the scattering of the impurity off the condensate. It has been shown \cite{Shchadilova:2016jzSM} that these terms are important for an accurate description of the physics of quantum impurities in ultracold gases. We start from Eq.~(\ref{eq:densitydensity}) but now we do not discard terms quadratic in the fluctuations fields; this gives rise to the following additional contribution
\begin{align}
\hat{H}_\text{imp-bos}^{(2)} &= \sum_{\mathbf{k}, \mathbf{q}} V_A(\mathbf{q}) \hat{\rho} (\mathbf{q}) \hat{A}^\dagger_{\mathbf{k} - \mathbf{q}} \hat{A}_\mathbf{k} + \sum_{\mathbf{k}, \mathbf{q}} V_B(\mathbf{q}) \hat{\rho} (\mathbf{q}) \hat{B}^\dagger_{\mathbf{k} - \mathbf{q}} \hat{B}_\mathbf{k} = \\
&= \sum_{\mathbf{k}, \mathbf{k}'} V_A(\mathbf{k}-\mathbf{k}') \hat{\rho} (\mathbf{k}-\mathbf{k}') \hat{A}^\dagger_{\mathbf{k}'} \hat{A}_\mathbf{k} + \sum_{\mathbf{k}, \mathbf{k}'} V_B(\mathbf{k}-\mathbf{k}') \hat{\rho} (\mathbf{k}-\mathbf{k}') \hat{B}^\dagger_{\mathbf{k}'} \hat{B}_\mathbf{k}
\end{align}
that can also be conveniently rewritten as
\beq
\hat{H}_\text{imp-bos}^{(2)} = \sum_{\mathbf{k}, \mathbf{k}'} V_A(\mathbf{k}+\mathbf{k}') \hat{\rho} (\mathbf{k}+\mathbf{k}') \hat{A}^\dagger_{-\mathbf{k}'} \hat{A}_\mathbf{k} + \sum_{\mathbf{k}, \mathbf{k}'} V_B(\mathbf{k}+\mathbf{k}') \hat{\rho} (\mathbf{k}+\mathbf{k}') \hat{B}^\dagger_{-\mathbf{k}'} \hat{B}_\mathbf{k} \; .
\label{eq:himpbos8}
\eeq
We recall that according to Eq.~(\ref{eq:brotation}) the fields transform under the Bogoliubov transformation in the following way
\beq
\begin{cases}
A_\mathbf{k} = M^{11}_\mathbf{k} a_\mathbf{k} + M^{12}_\mathbf{k} a^\dagger_{-\mathbf{k}} +  M^{13}_\mathbf{k} b_\mathbf{k}  + M^{14}_\mathbf{k} b^\dagger_{-\mathbf{k}} \\
A^\dagger_{-\mathbf{k}'} = M^{21}_{\mathbf{k}'} a_{\mathbf{k}'} + M^{22}_{\mathbf{k}'} a^\dagger_{-\mathbf{k}'} + M^{23}_{\mathbf{k}'} b_{\mathbf{k}'} + M^{24}_{\mathbf{k}'} b^\dagger_{-\mathbf{k}'} \\
B_\mathbf{k} = M^{31}_\mathbf{k} a_\mathbf{k} + M^{32}_\mathbf{k} a^\dagger_{-\mathbf{k}} +  M^{33}_\mathbf{k} b_\mathbf{k}  + M^{34}_\mathbf{k} b^\dagger_{-\mathbf{k}} \\
B^\dagger_{-\mathbf{k}'} = M^{41}_{\mathbf{k}'} a_{\mathbf{k}'} + M^{42}_{\mathbf{k}'} a^\dagger_{-\mathbf{k}'} +  M^{43}_{\mathbf{k}'} b_{\mathbf{k}'}  + M^{44}_{\mathbf{k}'} b^\dagger_{-\mathbf{k}'}
\end{cases}
\eeq
and let us assume that, due to rotational invariance $M^{ij}_\mathbf{k} = M^{ij}_{-\mathbf{k}}$. We then `split' the interaction term of Eq.~\ref{eq:himpbos8} as $\hat{H}_\text{imp-bos}^{(2)} = \hat{H}_\text{imp-bos}^{(2A)}  + \hat{H}_\text{imp-bos}^{(2B)}$ with
\beq
\begin{split}
\hat{H}_\text{imp-bos}^{(2A)} = \sum_{\mathbf{k}, \mathbf{k}'} V_A(\mathbf{k}+\mathbf{k}') \hat{\rho} (\mathbf{k}+\mathbf{k}') (M^{21}_{\mathbf{k}'} a_{\mathbf{k}'} + M^{22}_{\mathbf{k}'} a^\dagger_{-\mathbf{k}'} + M^{23}_{\mathbf{k}'} b_{\mathbf{k}'} + M^{24}_{\mathbf{k}'} b^\dagger_{-\mathbf{k}'}) \times \\
\times (M^{11}_\mathbf{k} a_\mathbf{k} + M^{12}_\mathbf{k} a^\dagger_{-\mathbf{k}} +  M^{13}_\mathbf{k} b_\mathbf{k}  + M^{14}_\mathbf{k} b^\dagger_{-\mathbf{k}})
\end{split}
\eeq
and
\beq
\begin{split}
\hat{H}_\text{imp-bos}^{(2B)} = \sum_{\mathbf{k}, \mathbf{k}'} V_B(\mathbf{k}+\mathbf{k}') \hat{\rho} (M^{41}_{\mathbf{k}'} a_{\mathbf{k}'} + M^{42}_{\mathbf{k}'} a^\dagger_{-\mathbf{k}'} +  M^{43}_{\mathbf{k}'} b_{\mathbf{k}'}  + M^{44}_{\mathbf{k}'} b^\dagger_{-\mathbf{k}'}) \times \\
\times (M^{31}_\mathbf{k} a_\mathbf{k} + M^{32}_\mathbf{k} a^\dagger_{-\mathbf{k}} +  M^{33}_\mathbf{k} b_\mathbf{k}  + M^{34}_\mathbf{k} b^\dagger_{-\mathbf{k}})
\end{split} \; .
\eeq
An alternative way of writing the extended interaction term is in matrix form, with
\beq
H^{(2A)}_\text{imp-bos} = \sum_{\mathbf{k}, \mathbf{k}'} V_A(\mathbf{k}+\mathbf{k}') \hat{\rho} (\mathbf{k}+\mathbf{k}') \begin{pmatrix}
\hat{a}_{\mathbf{k}'} \\
\hat{a}^\dagger_{-\mathbf{k}'} \\
\hat{b}_{\mathbf{k}'} \\
\hat{b}^\dagger_{-\mathbf{k}'} \\
\end{pmatrix}^T \begin{pmatrix}
 M^{11}_{\mathbf{k}} M^{21}_{\mathbf{k}'} &
 M^{12}_{\mathbf{k}} M^{21}_{\mathbf{k}'} &
 M^{13}_{\mathbf{k}} M^{21}_{\mathbf{k}'} &
 M^{14}_{\mathbf{k}} M^{21}_{\mathbf{k}'} \\

 M^{11}_{\mathbf{k}} M^{22}_{\mathbf{k}'} &
 M^{12}_{\mathbf{k}} M^{22}_{\mathbf{k}'} &
 M^{13}_{\mathbf{k}} M^{22}_{\mathbf{k}'} &
 M^{14}_{\mathbf{k}} M^{22}_{\mathbf{k}'} \\

 M^{11}_{\mathbf{k}} M^{23}_{\mathbf{k}'} &
 M^{12}_{\mathbf{k}} M^{23}_{\mathbf{k}'} &
 M^{13}_{\mathbf{k}} M^{23}_{\mathbf{k}'} &
 M^{14}_{\mathbf{k}} M^{23}_{\mathbf{k}'} \\

 M^{11}_{\mathbf{k}} M^{24}_{\mathbf{k}'} &
 M^{12}_{\mathbf{k}} M^{24}_{\mathbf{k}'} &
 M^{13}_{\mathbf{k}} M^{24}_{\mathbf{k}'} &
 M^{14}_{\mathbf{k}} M^{24}_{\mathbf{k}'}
\end{pmatrix} \begin{pmatrix}
\hat{a}_\mathbf{k} \\
\hat{a}^\dagger_{-\mathbf{k}} \\
\hat{b}_\mathbf{k} \\
\hat{b}^\dagger_{-\mathbf{k}} \\
\end{pmatrix}
\label{eq:matrixh2a}
\eeq
and
\beq
H^{(2B)}_\text{imp-bos} = \sum_{\mathbf{k}, \mathbf{k}'} V_B(\mathbf{k}+\mathbf{k}') \hat{\rho} (\mathbf{k}+\mathbf{k}') \begin{pmatrix}
\hat{a}_{\mathbf{k}'} \\
\hat{a}^\dagger_{-\mathbf{k}'} \\
\hat{b}_{\mathbf{k}'} \\
\hat{b}^\dagger_{-\mathbf{k}'} \\
\end{pmatrix}^T \begin{pmatrix}
 M^{31}_{\mathbf{k}} M^{41}_{\mathbf{k}'} &
 M^{32}_{\mathbf{k}} M^{41}_{\mathbf{k}'} &
 M^{33}_{\mathbf{k}} M^{41}_{\mathbf{k}'} &
 M^{34}_{\mathbf{k}} M^{41}_{\mathbf{k}'} \\

 M^{31}_{\mathbf{k}} M^{42}_{\mathbf{k}'} &
 M^{32}_{\mathbf{k}} M^{42}_{\mathbf{k}'} &
 M^{33}_{\mathbf{k}} M^{42}_{\mathbf{k}'} &
 M^{34}_{\mathbf{k}} M^{42}_{\mathbf{k}'} \\

 M^{31}_{\mathbf{k}} M^{43}_{\mathbf{k}'} &
 M^{32}_{\mathbf{k}} M^{43}_{\mathbf{k}'} &
 M^{33}_{\mathbf{k}} M^{43}_{\mathbf{k}'} &
 M^{34}_{\mathbf{k}} M^{43}_{\mathbf{k}'} \\
 
 M^{31}_{\mathbf{k}} M^{44}_{\mathbf{k}'} &
 M^{32}_{\mathbf{k}} M^{44}_{\mathbf{k}'} &
 M^{33}_{\mathbf{k}} M^{44}_{\mathbf{k}'} &
 M^{34}_{\mathbf{k}} M^{44}_{\mathbf{k}'} \\
\end{pmatrix} \begin{pmatrix}
\hat{a}_\mathbf{k} \\
\hat{a}^\dagger_{-\mathbf{k}} \\
\hat{b}_\mathbf{k} \\
\hat{b}^\dagger_{-\mathbf{k}} \\
\end{pmatrix}.
\label{eq:matrixh2b}
\eeq
which can easily be rewritten in compact form by introducing a spinor-like object $\Psi (\mathbf{k'})=(a_\mathbf{k} \ a^\dagger_{-\mathbf{k}} \ b_\mathbf{k} \ b^\dagger_{-\mathbf{k}} )^T$, as done in the main text.

\section{Energy functional of a heteronuclear self-bound droplet}
\label{sec:4impuritydroplet}

To study the effect of an impurity in the droplet phase we assume that the two components are described by a complex field $\phi_i(\mathbf{r})$ with the associated energy functional
\beq
E_\text{bb} = \int \mathrm{d}^3 r 
\left(\sum_{i=1,2}
\frac{\hbar^2|\nabla \phi_i|^2}{2m_i} 
+\frac{g_{ii}}{2}|\phi_i|^4
\right)+
g_{12}|\phi_1|^2|\phi_2|^2 + \frac{8}{15\pi^2\hbar^3}
\left(
m_1^\frac{3}{5}g_{11}|\phi_1|^2+
m_2^\frac{3}{5}g_{22}|\phi_2|^2
\right)^\frac{5}{2}.
\label{eq:droplet}
\eeq
The last term in Eq.~(\ref{eq:droplet}) is the
beyond mean-field interaction for a general two-component mixture \cite{Minardi:2019efSM}.
To probe the equilibrium properties, we set the relative number of particles in each of the two components to satisfy the constraint
$\frac{N_1}{N_2} = \sqrt{\frac{g_{22}}{g_{11}}}$ \cite{Petrov:2015khSM}.
Introducing the mass ratio $z=\frac{m_2}{m_1}$, the scaled mass $m^*=m \frac{1+\sqrt{z a_{11}/a_{22}}}{1+\sqrt{a_{11}/z a_{22}}}$, and the coupling constant ratio $\alpha=\sqrt{\frac{g_{11}}{g_{22}}} = z^{1/2}\sqrt{a_{11}/a_{22}}$ we can rewrite the 
densities of each component as follows
\beq
|\phi_1|^2 = \frac{1}{1+\alpha} |\phi|^2,\ \ \
|\phi_2|^2 = \frac{\alpha}{1+\alpha} |\phi|^2,
\eeq
where $|\phi(\mathbf{r})|^2=|\phi_1(\mathbf{r})|^2+
|\phi_2(\mathbf{r})|^2$.
Using the definition $\delta g = g_{12}+\sqrt{g_{11}g_{22}}$ we can rewrite the energy functional $E_\text{bb}$ as
\beq
E_\text{bb}[\phi] = \int \mathrm{d}^3 r \frac{\hbar^2|\nabla \phi|^2}{2m^*} +\delta g \frac{\alpha}{(1+\alpha)^2}
|\phi|^4 + 
\frac{8 m_1^\frac{3}{2}}{15\pi^2\hbar^3}
\left(\frac{g_{11}|\phi|^2}{1+\alpha}\right)^\frac{5}{2}
\left(1+\frac{z^\frac{3}{2}}{\alpha^\frac{1}{2}} \right)^\frac{5}{2}.
\label{eq:en_droplet_rescaled}
\eeq

Scaling lengths by $a_{11}$ and energies by $E_1 = \frac{\hbar^2}{m_1 a_{11}^2}$,
we can write a generalized Gross-Pitaevskii equation for the self-bound droplet
\beq
\begin{array}{ccl}
\mathrm{i} \frac{\partial \phi}{\partial t}   & =  &
\left( -\frac{\nabla^2}{2 m^*/m_1} +
g_{MF}|\phi|^2 + 
g_{LHY}|\phi|^3 \right)\phi(\mathbf{r},t).
\end{array}
\label{GPEd}
\eeq
where we introduced the effective couplings
\beq
\left\{
\begin{array}{ccl}
g_{MF} &=& 4\pi\frac{\alpha}{(1+\alpha)^2}
\left(\frac{2}{\alpha} + \frac{a_{12}}{a_{11}}\frac{1+z}{z} \right)\\ \\
g_{LHY} & = & \frac{128\sqrt{\pi}}{3}
\left( 
\frac{1+z^\frac{1}{10}\sqrt{a_{22}/a_{11}}}{1+\sqrt{z a_{11}/a_{22}}}
\right)^\frac{5}{2}.
\end{array}
\right.
\eeq
We solve Eq.~(\ref{GPEd}) to obtain, e.g., the density profiles of Figs.(\ref{fig:three}) and Fig.~(\ref{fig:four})d
(yellow shaded areas).
In Fig.~\ref{fig:six}
we plot the energy of the droplet as a function of the number of particles $N_D$ in the droplet (a), and as a function of the magnetic field $B$ (b).

\begin{figure*}[t]
\centering
    \includegraphics[width=.995\linewidth]{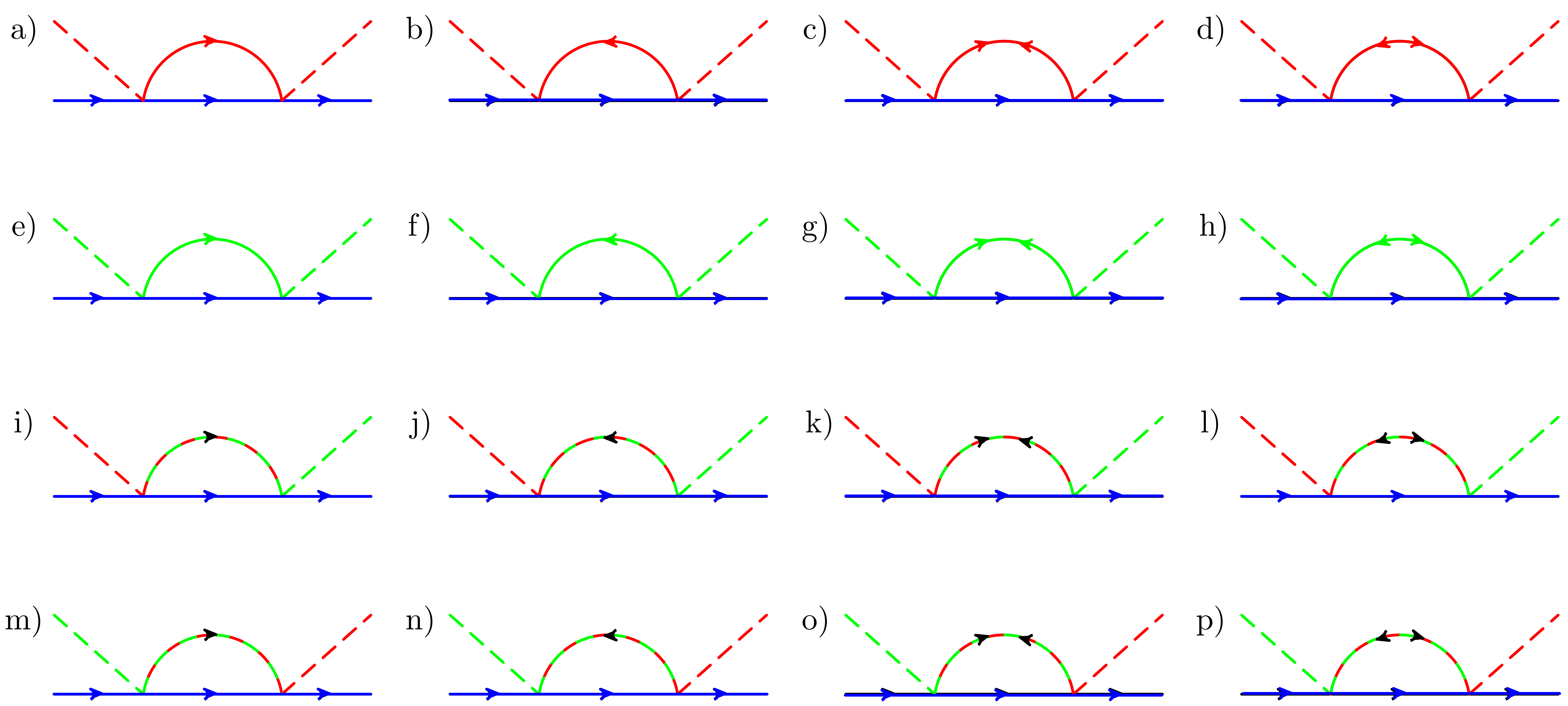}
\caption{Feynman diagrams of the interaction between the two-component BEC (red and green lines) and the impurity (blue lines). The arrows denote normal and anomalous propagators following the convention in Ref.~\cite{FetterWalecka}.}
\label{fig:five}
\end{figure*}

\section{Beyond-mean field impurity-BEC energy $\mathscr{E}_\text{BMF}(\mathbf{r})$}

The mean-field impurity-droplet potential reads
\beq
V_\text{eff}^\text{mf}(\mathbf{r}) = |\phi(\mathbf{r})|^2 
\left(\frac{1}{1+\alpha}\,g_{I1} +\frac{\alpha}{1+\alpha}\, g_{I2} \right). 
\label{eq:VeffMF}
\eeq
Introducing the effective coupling 
\beq
g_{ID} = \frac{2\pi}{1+\alpha}
\left(\frac{2a_{I1}}{a_{11} + \frac{\alpha a_{I2}}{a_{11}}\frac{1+z}{z}}\right)
\eeq
one obtains the first term of the impurity-droplet potential of Eq.~(\ref{Eimp}) of the main text.
We now compute the correction to the mean-field impurity-droplet energy in the perturbative limit for small $a_{Ii}/\xi_i$, ($i=1,2$) where $\xi_i=1/\sqrt{8\pi n_i a_{ii}}$ is the healing length of the $i$-th component of the BEC.
We focus on the second-order correction which can be derived using the generalized Fr\"ohlich Hamiltonian in Eq.~(\ref{frolich}). Within this approximation, the interaction Hamiltonian reads
\beq
\begin{array}{ccl}
\hat{H}_\text{imp-bos} &=&
\sqrt{n_A} \sum_{{\bf q} ,{\bf k}} V_A({\bf q}) \ e^{\mathrm{i} \mathbf{q} \cdot \hat{\mathbf{R}}}\, 
(A_{\bf q} + A^\dagger_{\bf -q}) +
\sqrt{n_B} \sum_{{\bf q} ,{\bf k}} V_B({\bf q}) e^{\mathrm{i} \mathbf{q} \cdot \hat{\mathbf{R}}}\, 
(B_{\bf q} + B^\dagger_{\bf -q})\\ \\
&=& \sqrt{n_A} \sum_{{\bf q} ,{\bf k}} V_A({\bf q}) \ e^{\mathrm{i} \mathbf{q} \cdot \hat{\mathbf{R}}}
\left( 
(M_{\bf q}^{11}+M_{\bf q}^{21}) \hat{a}_{\bf q} + 
(M_{\bf q}^{12}+M_{\bf q}^{22}) \hat{a}^\dagger_{\bf -q} +
(M_{\bf q}^{13}+M_{\bf q}^{23}) \hat{b}_{\bf q} + 
(M_{\bf q}^{14}+M_{\bf q}^{24}) \hat{b}^\dagger_{\bf -q}
\right)+ \\ \\
&+& \sqrt{n_B} \sum_{{\bf q} ,{\bf k}} V_B({\bf q}) \ e^{\mathrm{i} \mathbf{q} \cdot \hat{\mathbf{R}}}
\left( 
(M_{\bf q}^{31}+M_{\bf q}^{41}) \hat{a}_{\bf q} + 
(M_{\bf q}^{32}+M_{\bf q}^{42}) \hat{a}^\dagger_{\bf -q} +
(M_{\bf q}^{33}+M_{\bf q}^{43}) \hat{b}_{\bf q} + 
(M_{\bf q}^{34}+M_{\bf q}^{44}) \hat{b}^\dagger_{\bf -q}
\right).
\end{array}
\eeq
We then apply perturbation theory to find the correction to the ground state energy. The first-order correction 
$\Delta E^{(1)}$ vanishes.
The second-order correction reads
\beq
\begin{array}{clc}
\Delta E^{(2)}&=& 
n\frac{1}{1+\alpha} 
\left(\frac{2\pi \hbar^2 a_{1I}}{\mu_{1I}}\right)^2 
\int \frac{\mathrm{d}^3 k}{(2\pi)^3}
\left(-
\frac{|(M_\mathbf{k}^{12}+M_\mathbf{k}^{22})+\sqrt{\alpha}\frac{a_{2I}}{a_{1I}} \frac{\mu_{1I}}{\mu_{2I}} (M_\mathbf{k}^{32}+M_\mathbf{k}^{42})|^2}{\omega_k^{(A)}}
+\frac{2\mu_{1I}}{\hbar^2 k^2}
\right)\\ \\
&+&
n\frac{\alpha}{1+\alpha} 
\left(\frac{2\pi \hbar^2 a_{2I}}{\mu_{2I}}\right)^2 
\int \frac{\mathrm{d}^3 k}{(2\pi)^3}
\left(-
\frac{|(M_\mathbf{k}^{34}+M_\mathbf{k}^{44})+\left(\sqrt{\alpha}\frac{a_{2I}}{a_{1I}} \frac{\mu_{1I}}{\mu_{2I}}\right)^{-1} (M_\mathbf{k}^{14}+M_\mathbf{k}^{24})|^2}{\omega_k^{(B)}}
+\frac{2\mu_{2I}}{\hbar^2 k^2}
\right).
\end{array}
\label{deltaE2}
\eeq
The last term in the two integrands acts as a regularizing term for the pair integrals and it has the same form as in Lippmann-Schwinger equation for each one of the two components. The second-order energy shift can be equivalently expressed as the sum of Feynman diagrams shown in Fig.~(\ref{fig:five}).

Finally, Eq.~(\ref{deltaE2}) can be recast as in Eq.~(\ref{Ebmf}) of the main text upon introducing the healing lengths of the each component $\xi_i = (8\pi n_i a_{ii})^{-1/2}$, $i=1,2$. 
With the help of the substitutions 
$\frac{\hbar^2 k^2}{2 m_i}=K^2 n_i \frac{4\pi\hbar^2\,a_{ii}}{2m_i}$, $i=1,2$ we obtain
\beq
\mathscr{E}_\text{BMF} = \frac{1}{1+\alpha} \left( \frac{2 \pi \hbar^2 \xi_1\, n}{\mu_{I1}} \right) \left( \frac{a_{I1}}{\xi_1} \right)^2 \frac{m_1}{\mu_{I1}} I_1
+ \frac{\alpha}{1+\alpha} \left( \frac{2 \pi \hbar^2 \xi_2 \, n}{\mu_{I2}} \right) \left( \frac{a_{I2}}{\xi_2} \right)^2 \frac{m_2}{\mu_{I2}} I_2,
\label{eq:EIBMF}
\eeq
where we introduced the dimensionless integrals
\beq
\begin{array}{ccl}
I_1 &=& \frac{2}{\pi}
\int \mathrm{d}^3 K
\left(-
\frac{|(M_\mathbf{K}^{12}+M_\mathbf{K}^{22})+\sqrt{\alpha}\frac{a_{2I}}{a_{1I}} \frac{\mu_{1I}}{\mu_{2I}} (M_\mathbf{K}^{32}+M_\mathbf{K}^{42})|^2}{\omega_K^{(A)}}
+\frac{\mu_{1I}}{m_1}\frac{1}{K^2}
\right)
\\ \\
I_2 &=& \frac{2}{\pi}
\int  \mathrm{d}^3 K
\left(-
\frac{|(M_\mathbf{K}^{34}+M_\mathbf{K}^{44})+\left(\sqrt{\alpha}\frac{a_{2I}}{a_{1I}} \frac{\mu_{1I}}{\mu_{2I}}\right)^{-1} (M_\mathbf{K}^{14}+M_\mathbf{K}^{24})|^2}{\omega_K^{(B)}}
+\frac{\mu_{2I}}{m_2}\frac{1}{K^2}
\right).
\end{array}
\eeq

We employ local-density approximation to account for the space-dependent condensate profiles, leading to a \textit{generalized} Lee-Huang-Yang type of beyond-mean field energy shift
for the impurity.
The sum of the contributions of Eq.~(\ref{eq:VeffMF}) and Eq.~(\ref{eq:EIBMF}) 
concludes the derivation of the effective potential in Eq.~(\ref{Eimp}) of the main
text

\beq
V_\text{eff}(\mathbf{r})
= g_{ID}|\phi (\mathbf{r})|^2 + \mathscr{E}_\text{BMF} (\mathbf{r})
\eeq

\begin{figure*}[t]
\centering
    \includegraphics[width=.995\linewidth]{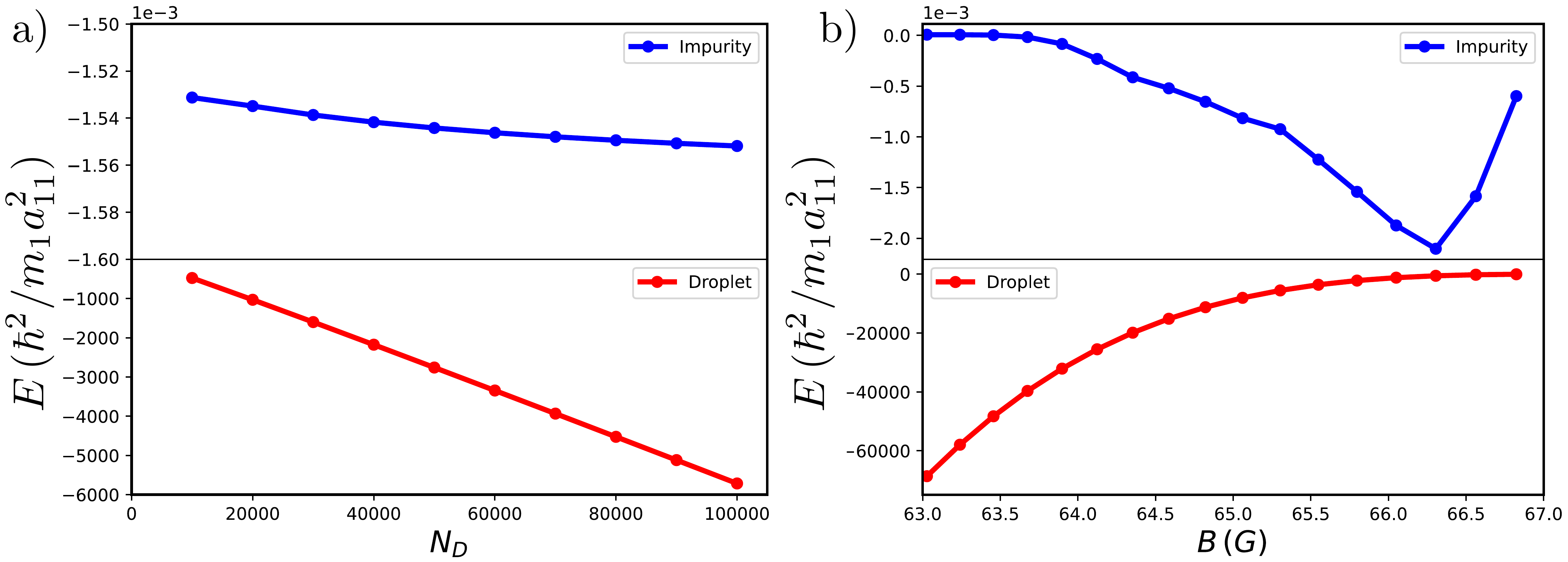}
\caption{Energies $E_I$ of the impurity (blue) and $E_D$ of the droplet (red) in units of $\hbar^2/m_1 a_{11}^2$ as a function of a) the number of particles in the droplet $N_D$ and b) the magnetic field $B\,$(G). In a) we set $B=65.8$ G and in b) $N_D=4\times 10^4$.}
\label{fig:six}
\end{figure*}

In Fig.~\ref{fig:six} we plot the energy of the impurity as a function of the number of particles in the droplet (a), and as a function of the magnetic field (b). 
We observe that the impurity energy is a non-monotonic function of the magnetic field.
Finally, we notice that, due to the large energy difference between the droplet and the impurity, we neglect the backaction of the impurity on the droplet. 

\end{document}